\documentclass[aps,prapplied,twocolumn,superscriptaddress]{revtex4-2}

\usepackage{graphicx}
\usepackage{dcolumn}
\usepackage{bm}

\begin{document}



\title{Tracking and fast imaging of a moving object via Fourier modulation}

\author{Shijian Li}
\affiliation{School of Integrated Circuits and Electronics, Beijing Institute of Technology, Beijing 100081, China}
\author{Xu-Ri Yao}
\email{yaoxuri@bit.edu.cn}
\affiliation{Center for Quantum Technology Research and Key Laboratory of Advanced Optoelectronic Quantum Architecture and Measurements (MOE), School of Physics, Beijing Institute of Technology, Beijing 100081, China}

\author{Wei Zhang}
\affiliation{Center for Quantum Technology Research and Key Laboratory of Advanced Optoelectronic Quantum Architecture and Measurements (MOE), School of Physics, Beijing Institute of Technology, Beijing 100081, China}
\author{Yeliang Wang}
\affiliation{School of Integrated Circuits and Electronics, Beijing Institute of Technology, Beijing 100081, China}
\author{Qing Zhao}
\email{qzhaoyuping@bit.edu.cn}
\affiliation{Center for Quantum Technology Research and Key Laboratory of Advanced Optoelectronic Quantum Architecture and Measurements (MOE), School of Physics, Beijing Institute of Technology, Beijing 100081, China}

\date{\today}

\begin{abstract}
Recently, several single-pixel imaging (SPI) schemes have emerged for imaging fast-moving objects and have shown dramatic results. However, fast image reconstruction of a moving object with high quality is still challenging for SPI, thereby limiting its practical application. In this paper, we present a simultaneous tracking and imaging method that incorporates position encoding and spatial information encoding through Fourier patterns. The utilization of Fourier patterns with specific spatial frequencies ensures robust and accurate object localization. By exploiting the properties of the Fourier transforms, our method achieves a remarkable reduction in time complexity while significantly enhancing image quality. Furthermore, we introduce an optimized sampling strategy specifically designed for small moving objects, significantly reducing the required dwell time for imaging. The proposed method provides a practical solution for real-time tracking, imaging, and edge detection of moving objects, underscoring its considerable potential for diverse applications. 
\end{abstract}

\maketitle


\section{Introduction}
The precise tracking and imaging of fast-moving objects hold tremendous promise for various applications, including robot navigation~\cite{adamkiewicz2022vision}, cell sorting~\cite{abj3013}, and unmanned aerial vehicle detection~\cite{jiang2021anti}. Single-pixel imaging (SPI), an emerging imaging technology, has several key advantages, such as a widely applicable spectrum and high detection sensitivity, making it an appealing tool with applications across various fields~\cite{howland2011photon,phillips2017adaptive,gibson2017real,bian2018experimental,edgar2019principles,stantchev2020real,jiang20222,liu2023optical}. SPI usually requires a large amount of time to obtain numerous samples and reconstruction algorithms are executed to image a stationary scene with high quality. When the object moves at high speed, using SPI methods directly can result in image degradation or blurring due to insufficient measurements within a single motion frame.

There are two primary approaches to achieve the SPI of moving objects. One involves enhancing the modulation speed, whereas the other focuses on estimating the motion parameters. An increased modulation speed can be achieved through various methods, such as LED-based illumination~\cite{xu20181000}, mechanical translation masks~\cite{jiang2020imaging}, spinning masks~\cite{hahamovich2021single,jiang2021single}, or swept aggregate patterns~\cite{kilcullen2022compressed}. Despite achieving higher modulation rates than the commonly used digital micro-mirror device (DMD), these methods typically achieve spatial resolutions of up to $101 \times 103 $ pixels~\cite{hahamovich2021single,kilcullen2022compressed}, presenting challenges for enhancing the spatial resolution. Moreover, they necessitate additional algorithms to process the obtained sequence of motion frames to achieve object tracking. Motion parameter estimation methods primarily address translational objects and employ a range of techniques, including algorithm estimation~\cite{zhang2013improving,jiao2019motion}, the projection-slice theorem~\cite{jiang2017adaptive,shi2019fast,yang2022image}, low-resolution image correlation calculations~\cite{sun2019gradual,monin2021single,wu2021fast}, Fourier patterns~\cite{zhang2019image,dan2022motion,li2023single}, geometric moment patterns~\cite{zha2021single,xiao2022single,zhang2023mask}, two-dimensional projective patterns~\cite{Yang2022Anti-motion}, laterally shifting patterns~\cite{sun2022simultaneously}, and four-quadrant detectors~\cite{du2023information} to measure object displacement. Among these methods, the Fourier pattern-based approaches~\cite{zhang2019image,dan2022motion,li2023single} excel in reliably capturing relative displacements at high frame rates but may not provide absolute positions. Geometric moment-based tracking methods~\cite{zha2021single,xiao2022single,zhang2023mask} generally offer the highest tracking frame rates, but they exhibit lower noise robustness than to Fourier pattern-based methods~\cite{li2023single}. Motion compensation is then applied to the imaging patterns, and object reconstruction is achieved via ghost imaging algorithms~\cite{bromberg2009ghost} or compressed sensing algorithms~\cite{li2013efficient}. Moreover, most of the imaging patterns used in these approaches are either random patterns or Hadamard patterns. After motion compensation, using fast inverse transformations for image reconstruction may not be feasible. Consequently, a time-consuming nonlinear iterative algorithm is essential for these methods to achieve high-quality reconstruction, making it difficult to track and do fast imaging simultaneously.

In this study, we present an approach for achieving high-frame-rate tracking, fast imaging, and edge detection of a moving object. The proposed approach employs Fourier patterns for both absolute position tracking and object imaging. First, six tracking patterns with specific spatial frequencies are utilized for each motion frame to ensure robust and accurate absolute position tracking, overcoming the limitations of traditional Fourier methods~\cite{zhang2019image,dan2022motion,li2023single}, which can obtain only relative positions. Then, by employing phase correction and the inverse fast Fourier transform (iFFT) on the acquired Fourier spectrum from the imaging patterns, we achieve fast and high-quality imaging as well as edge detection concurrently. We demonstrate this experimentally by fast-tracking and imaging a real moving object. Additionally, we propose an optimized sampling strategy for small objects and validate it by monitoring a fast-moving object. The proposed method has potential for application in scenarios requiring real-time object detection.
\section{Theory and analysis}
As demonstrated in Fourier single-pixel imaging (FSI)~\cite{zhang2015single,zhang2017fast,deng2019fourier,liu2023fourier}, the spatial characteristics of an object are encoded through a spatial light modulator (SLM) utilizing Fourier basis patterns. The series of modulated total light intensities are subsequently detected by a single-pixel detector. A Fourier basis pattern $P(x, y)$ can be represented by its corresponding spatial frequency pair $(k_x, k_y)$ and corresponding initial phase $\varphi_0$: 
\begin{equation}
	\label{eq:fs}
	P\left(x, y \mid k_{x}, k_{y}, \varphi_{0}\right)=a_0+b_0\cos \left[2 \pi\left(k_{x} x+k_{y} y\right)+\varphi_{0}\right],
\end{equation}
where $a_0$, $b_0$, and $(x, y)$ represent the average intensity, contrast, and two-dimensional Cartesian coordinates of the pattern, respectively. The modulated total light intensity $I_\varphi(k_{x}, k_{y})$ can be obtained via such Fourier basis patterns:
\begin{equation}
	\label{eq:fimg}
	I_\varphi(k_{x}, k_{y}) =I_b+\alpha\sum_{x} \sum_{y}O(x, y)P(x, y \mid k_{x}, k_{y}, \varphi),
\end{equation}
where $I_b$, $\alpha$, and $O(x, y)$ represent the intensity of the background noise, a constant, and the object image, respectively. The Fourier coefficients of the object image are obtained by these intensities. In the four-step phase-shifting method, four Fourier patterns, each with the same spatial frequency but different initial phases (0, $\pi/2$, $\pi$, and $3\pi/2$), are used to capture a Fourier coefficient. The Fourier spectrum $\tilde{I}(k_{x}, k_{y})$ is given by 
\begin{eqnarray}
   &&\tilde{I}(k_{x}, k_{y})\nonumber\\
   &&=I_{0}(k_{x}, k_{y})-I_{\pi}(k_{x}, k_{y}) 
	+j[I_{\pi/2}(k_{x}, k_{y})-I_{3\pi/2}(k_{x}, k_{y})] \nonumber\\
 &&=2b_0\alpha\mathcal{F}\{O(x,y)\},
\end{eqnarray}
where $\mathcal{F}$ represents the Fourier transform. Similarly, the three-step phase-shifting method involves three Fourier patterns with the same spatial frequency but initial phases of 0, $2\pi/3$, and $4\pi/3$. The three-step phase-shifting method offers high efficiency, whereas the four-step phase-shifting method provides better noise resistance\cite{zhang2017fast}. The Fourier spectrum can be sampled via either three-step or four-step phase-shifting methods as needed.  Once the Fourier spectrum of the object is obtained, the image can be reconstructed through the inverse Fourier transform. 

According to the translation property of the Fourier transform, the displacements $\Delta x$ and $\Delta y$ of a translational object in the spatial domain will result in a phase shift in the Fourier domain, which can be expressed as
\begin{eqnarray}
	&&O(x+\Delta x, y+\Delta y) \nonumber\\
 &&=\mathcal{F}^{-1}\{\tilde{I}(k_{x}, k_{y}) 
	\exp [2 \pi j (k_{x} \Delta x+k_{y} \Delta y)]\},
\end{eqnarray}
where $\mathcal{F}^{-1}$ represents the inverse Fourier transform. The displacements $\Delta x$ and $\Delta y$ can be restored by measuring the phase shift term $\varphi=2\pi(k_x \Delta x+k_y \Delta y)$. To avoid motion blur, phase changes due to target motion should be corrected, so the position of the object in each motion frame needs to be determined.

\begin{figure}
\centering\includegraphics[width=8.6 cm]{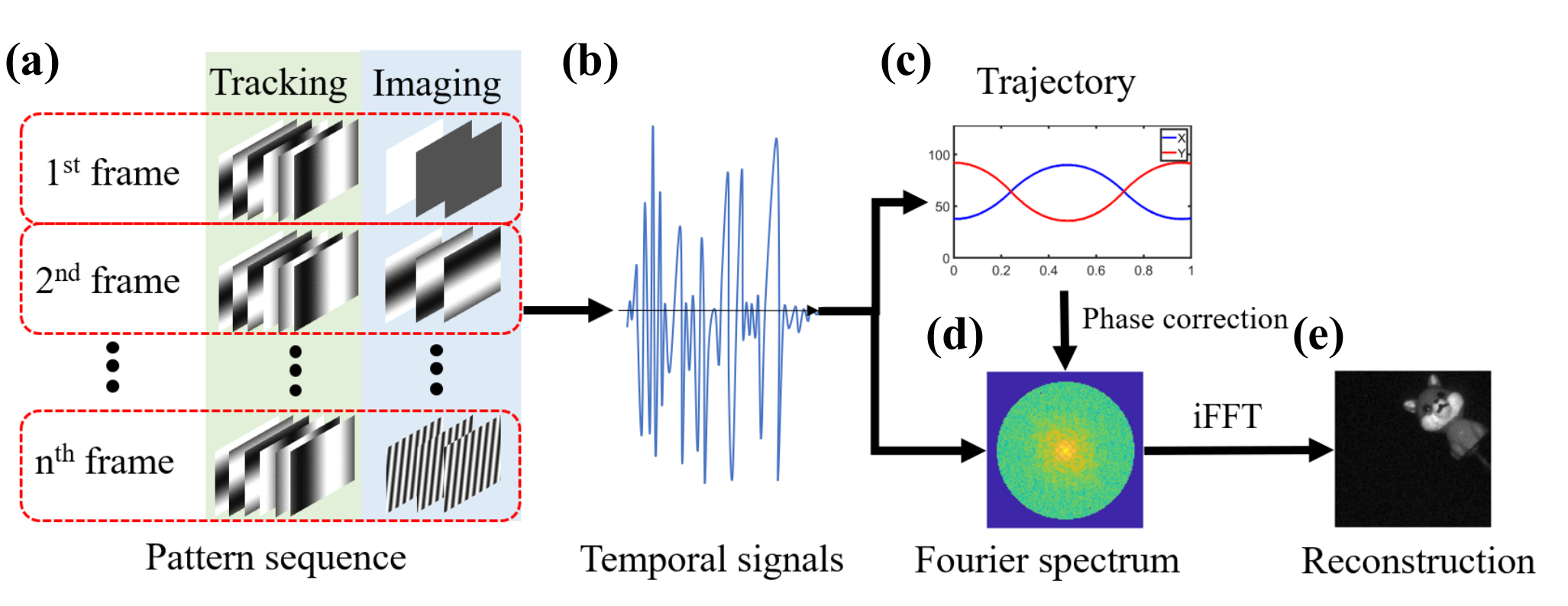}
 \caption{Schematic of the proposed method. (a) Pattern sequence for tracking and imaging. (b) The temporal signals detected via a bucket detector. (c) The calculated trajectories in two directions from the temporal signals. (d) The phase-corrected Fourier spectrum of the object. (e) The image restored via the inverse fast Fourier transform (iFFT).}
 \label{fgr:fig1}
\end{figure}
A schematic of the proposed method is depicted in Fig.~\ref{fgr:fig1}. The modulation patterns are divided into two components: tracking patterns and imaging patterns, as illustrated in Fig.~\ref{fgr:fig1}(a). The tracking patterns remain consistent across all motion frames, comprising six patterns generated via the three-step phase-shifting method for fast-tracking. These patterns correspond to spatial frequencies $(k_x=1/M, k_y=0)$ and $(k_x=0, k_y=1/N)$, where $M$ and $N$ represent the number of pixels in the $X$ and $Y$ directions, respectively. The two Fourier coefficients $\tilde{I_i}(1/M,0)$ and $\tilde{I_i}(0,1/N)$ of each motion frame can be obtained via such patterns. Therefore, the displacements in one direction, for example, the X-direction, between the $i^{th}$ frame and the initial frame can be calculated as follows:
\begin{eqnarray}
	\Delta x_i  &=& \frac{1}{2\pi (1/M)} \{ \arg [\tilde{I_i}(1/M, 0)]-\arg[\tilde{I_1}(1/M,0)]\}  \nonumber\\
	&=& \frac{\theta_{xi}}{2\pi (1/M)}-\frac{\theta_{x1}}{2\pi (1/M)} \nonumber\\
	&=&x_i-x_1,
\end{eqnarray}
where $\arg[]$ denotes the argument operation, $\tilde{I_1}\left(1/M, 0\right)$ represents the Fourier coefficient obtained at the initial frame, and $\tilde{I_i}(1/M,0)$ represents the Fourier coefficient obtained at the current motion frame. $\theta_{xi}$ and $\theta_{x1}$ are the corresponding phase angles of the Fourier coefficient. $x_i$ and $x_1$ are the coordinates obtained from the corresponding Fourier coefficients. Similarly, the relative displacement $\Delta y_i$ in the $Y$-direction at the $i^{th}$ frame, as well as the coordinate $y_i$, can also be determined as
\begin{eqnarray}
	\Delta y_i  &=& \frac{1}{2\pi (1/N)}\{ \arg [\tilde{I_i}(0,1/N)]-\arg[\tilde{I_1}(0,1/N)]\}  \nonumber\\
	&=& \frac{\theta_{yi}}{2\pi (1/N)}-\frac{\theta_{y1}}{2\pi (1/N)}  \nonumber\\
	&=&y_i-y_1.
\end{eqnarray}
The acquired coordinates ($x_i$, $y_i$), relative to the object's position, remain unchanged during translation. These coordinates can serve as feature points for object tracking. Therefore, the absolute position coordinates ($x_i$, $y_i$) of the object at the $i^{th}$ frame,  can be obtained as
\begin{eqnarray}
	\label{eq:cmpdx}
	x_i&=&\frac{\theta_{xi}}{2\pi (1/M)},\nonumber\\
	y_i&=&\frac{\theta_{yi}}{2\pi (1/N)},
\end{eqnarray}
where $\theta_{xi}$ and $\theta_{yi}$ are the corresponding phase angles of the Fourier coefficients $\tilde{I_i}(1/M,0)$ and $\tilde{I_i}(0,1/N)$, respectively.

The imaging patterns in each motion frame are composed of three (using the three-step phase-shifting method) or four (using the four-step phase-shifting method) Fourier patterns, each corresponding to a single spatial frequency. The spatial frequencies of the imaging patterns vary across different motion frames. They are chosen on the basis of spatial frequency position within the Fourier spectrum. The fundamental approach follows a circular order~\cite{zhang2017hadamard} and is referred to as the sequential sampling strategy. In scenarios where the field of view (FOV) has an even number of pixels in both directions, and where the object's size within the X and Y dimensions occupies less than half of it, we employ an alternative strategy. This strategy involves selecting imaging patterns at regular intervals within the Fourier spectrum, corresponding to spatial frequencies such as $k_x= 0/M, 2/M, 4/M,..., (M-2)/M$ and $ k_y = 0/N, 2/N, 4/N,..., (N-2)/N$. This approach is referred to as the interval sampling strategy. By using such a strategy, an image containing four copies of the object can be generated, one of which represents the actual object. The image with dimensions of $M/2 \times N/2$ pixels can subsequently be determined by the tracked coordinates of the object at its initial position. 

After tracking the trajectory of the object (Fig.~\ref{fgr:fig1}(c)) and subsequently calculating the displacement from the initial position, the phase variation of the Fourier coefficient caused by motion in the $i^{th}$ frame can be corrected via Eq.~(\ref{eq:corr}) to obtain the phase-corrected Fourier spectrum, as shown in Fig.~\ref{fgr:fig1}(d).
\begin{equation}
	\tilde{I}_{corr}\left(k_{x}, k_{y}\right)=\tilde{I_i}\left(k_{x}, k_{y}\right) \exp \left[-2 \pi j \left(k_{x} \Delta x_i+k_{y} \Delta y_i\right)\right].
	\label{eq:corr}
\end{equation} 
The object image can subsequently be extracted through iFFT from the corrected Fourier spectrum, as illustrated in Fig.~\ref{fgr:fig1}(e).

The existing edge detection algorithms detected in the Fourier domain~\cite{ren2018edge,wu2024edge} can be used to obtain real-time edge images of moving objects after a phase correction. This involves applying Sobel operators in the $X$ and $Y$ directions to detect edges. The Sobel operators $s_x$ and $s_y$ are defined as
\begin{eqnarray}
	s_x&&=\left[
	\begin{array}{r rr}
		-1 & \phantom{-}0 & \phantom{-}1\\
		-2 & \phantom{-}0 & \phantom{-}2 \\
		-1 & \phantom{-}0 &\phantom{-}1 \\
	\end{array}\right],\nonumber\\
	s_y&&=\left[
	\begin{array}{rrr}
		-1 & -2 & -1\\
		\phantom{-}0 & \phantom{-}0 & \phantom{-}0 \\
		\phantom{-}1 & \phantom{-}2 & \phantom{-}1 \\
	\end{array}\right].
\end{eqnarray}
These Sobel operators, each of size $3\times3$, are zero-padded to match the dimensions of the object image and are placed at the center of the zero-padding matrices to maintain the position of the edge image relative to the original image unchanged. The zero-padded matrices are denoted as $S_x$ and $S_y$, respectively. The resulting edge image, denoted as $E(x,y)$, is calculated as follows:
\begin{equation}
	E(x,y)=\sqrt{[\mathcal{F}^{-1}(\tilde{I}_{corr} \tilde{S}_x)]^2+[\mathcal{F}^{-1}(\tilde{I}_{corr}  \tilde{S}_y)]^2},
\end{equation}
Here, $\tilde{I}_{corr}$ represents the phase-corrected Fourier spectrum, whereas $\tilde{S_x}$ and $\tilde{S_y}$ correspond to the Fourier transforms of the Sobel operators in two different directions. 

\section{Results and Discussion}
\subsection{Simulation results}
\begin{figure}
	\centering\includegraphics[width=8.6 cm]{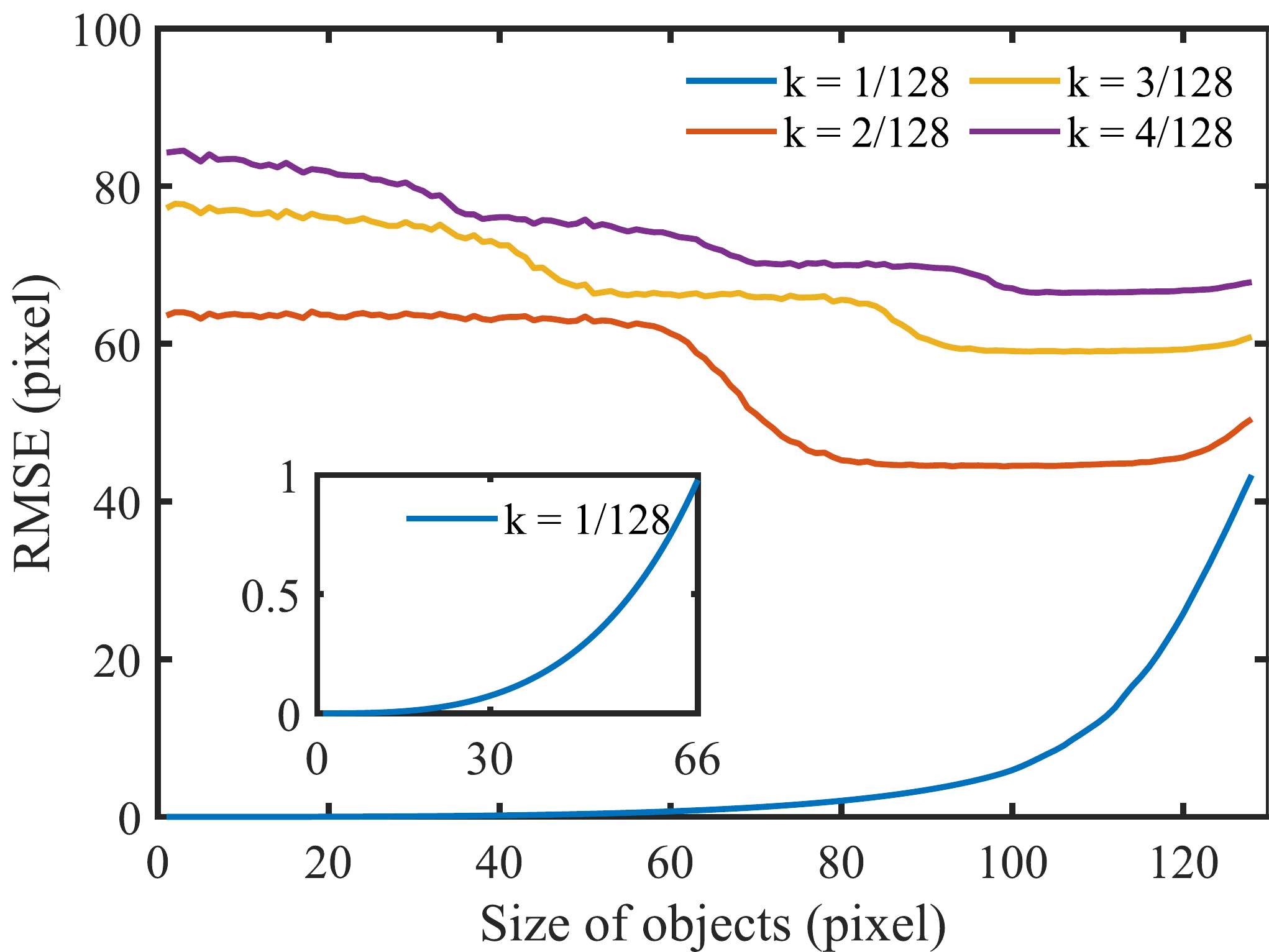}
	\caption{The root-mean-square errors (RMSEs) when patterns with different spatial frequencies are used for location.}
	\label{fgr:fig0}
\end{figure}

\begin{figure}
 \centering
 \includegraphics[width=8.6 cm]{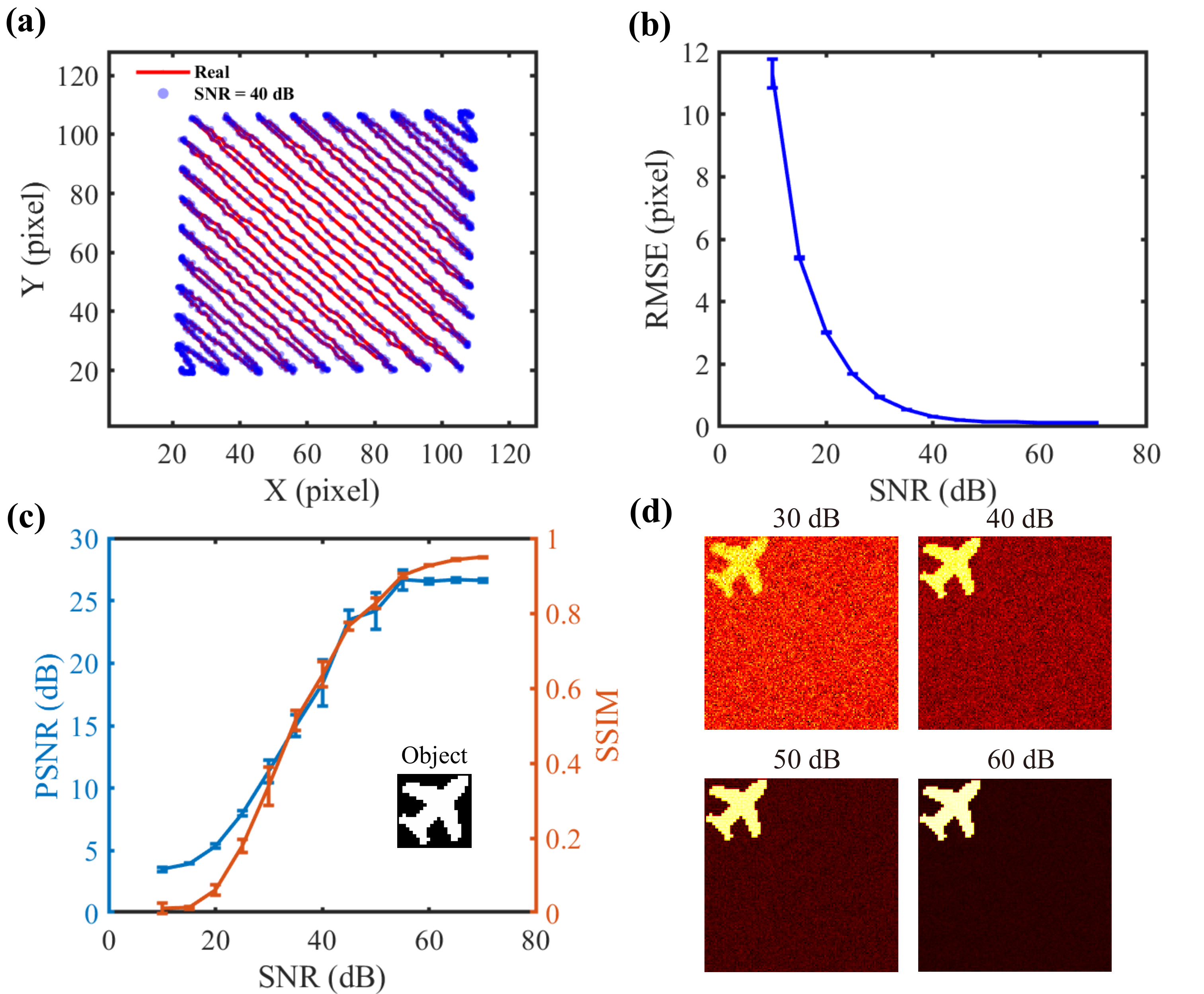}
 \caption{The simulation results of the proposed method under different noise levels. (a) Real trajectory and tracked trajectory using our method under a noise level of 40~$\mathrm{dB}$. (b) The RMSE of the tracking trajectory under different noise levels. (c) PSNR and SSIM curves of the reconstructed images under different noise levels. (d) The reconstructed images under four different noise levels. }
 \label{fgr:sim}
\end{figure}

We initially conducted a simulation to verify the ability of the proposed method to locate an object. Specifically, we employed 800 grayscale images from the DIV2K dataset~\cite{agustsson2017ntire}, scaling them to various dimensions and randomly positioning them as virtual objects within a $128 \times 128$-pixel FOV. We utilized Fourier patterns with four spatial frequencies of $1/128$, $2/128$, $3/128$, and $4/128$ for location. The root-mean-square error (RMSE) between the location coordinates and the image centroid coordinates was introduced to evaluate the accuracy of the calculated coordinates. We conducted ten locating simulations for objects of each size and averaged the errors. As depicted in Fig.~\ref{fgr:fig0}, our method, which employs patterns with a spatial frequency of $1/128$, locates simulated objects with smaller RMSEs across all sizes than the other three spatial frequency patterns. The calculated RMSEs of our method remain consistently below half of the object's size for all sizes of simulated objects, indicating that the calculated coordinates consistently fall within the boundaries of the objects. When the object size decreased below $66 \times 66$ pixels, the RMSEs of our method remained below one pixel. Given that the coordinates obtained relative to the object's position remain unaffected during translation, our method showcases its capability for absolute position tracking.
To study the influence of noise on the reconstructed image quality and tracking accuracy via the proposed method, we conducted another simulation to evaluate noise robustness by adding Gaussian white noise with different signal-to-noise ratios (SNRs). A virtual object, sized $43\times43$ pixels, was moved within an FOV of $128\times128$ pixels. A total of 8194 motion frames were simulated to fully sample the entire FOV. The tracking and imaging patterns were constructed via the three-step phase-shifting method, which requires nine patterns per motion frame (six for tracking and three for imaging). We also employed quantitative factors such as the peak signal-to-noise ratio (PSNR) and structural similarity (SSIM)~\cite{wang2004image} to evaluate the quality of the reconstructed images. The real trajectory and the tracked trajectory using our method under a noise level of 40~$\mathrm{dB}$ are shown in Fig.~\ref{fgr:sim}(a). We calculated the tracking RMSEs and the quantitative analysis curves of the reconstructed images under different noise levels, as shown in Fig.~\ref{fgr:sim}(b) and Fig.~\ref{fgr:sim}(c), respectively. Four sample images under four different noise levels are also shown in Fig.~\ref{fgr:sim}(d). It can be seen that noise significantly impacts image quality and tracking accuracy. When the SNR of the measurements is greater than 40~$\mathrm{dB}$, our method can track and image the object well.

\subsection{Experimental results}
\begin{figure}
 \centering\includegraphics[width=8.6 cm]{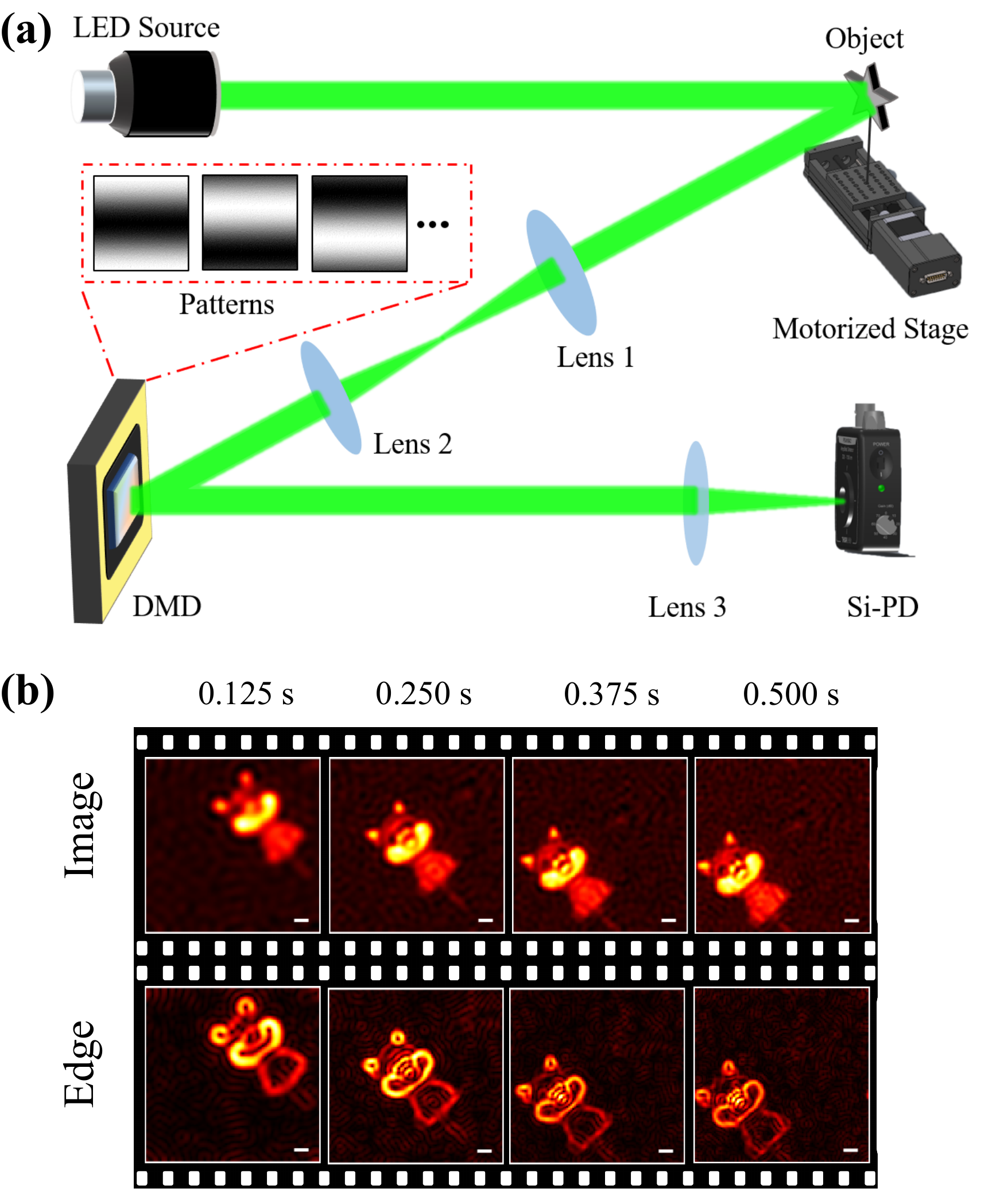}
 \caption{Schematic of the experimental setup and the images reconstructed at different times via our method. (a) The object is loaded on a motorized stage and imaged on digital micro-mirror device (DMD) following the light-emitting diode (LED) illumination. The modulated light is collected by a single-pixel silicon-based photodetector (Si-PD). (b) The images and edges at different times obtained via our method. All scale bars indicate 5~$\mathrm{mm}$. }
 \label{fgr:exp}
\end{figure}
As illustrated in Fig.~\ref{fgr:exp}(a), the experimental setup comprises several key components, including a high-power light-emitting diode (LED) source with a maximum output of 3.6 watts, a linear motorized stage (X-LSQ300B-E01, Zaber), a DMD (DLP7000, Texas Instruments), and a silicon-based photodetector (Si-PD, PDA100A2, Thorlabs). Following modulation by the DMD, the modulated light is collected by the Si-PD and converted into measurable values through the data acquisition board (USB-6341, National Instruments). The modulation patterns on the DMD have a pixel size of $256\times256$, where every $2\times2$ pixel is merged into a single superpixel. Consequently, the FOV has an image size of $128\times128$ pixels. The Fourier patterns employed in our method are binarized using the Sierra-Lite dithering algorithm~\cite{liang2019fast} with an upsampling ratio of 2.

We first experiment to achieve rapid tracking and imaging of a moving object. With the prior knowledge that the Fourier spectrum of any real-valued image is conjugated symmetric, a pair of conjugate Fourier coefficients can be obtained by sampling one single Fourier coefficient. To facilitate comparison with the conventional method using Hadamard patterns, the imaging patterns in this experiment were constructed via the four-step phase-shift method, which requires ten Fourier patterns for each motion frame. Among these, the former six Fourier patterns were tracking patterns, whereas the remaining four were constructed via the four-step phase-shifting method for imaging. We substituted the four Fourier patterns for imaging each motion frame with four differential Hadamard patterns as the conventional method for a fair comparison. The Hadamard patterns corresponding to different motion frames were sorted according to the total variation (TV) ordering method~\cite{yu2020super}. The conventional method necessitates compensating for imaging patterns by incorporating the relative displacement obtained through object tracking, thereby generating new measurement patterns. Therefore, the Hadamard patterns after motion compensation cannot use fast inverse transformation. This method, which uses Hadamard patterns, referred to as Had\_GI, employs ghost imaging algorithms~\cite{bromberg2009ghost} to swiftly restore the object's image. We also employed the pseudoinverse ghost imaging algorithm~\cite{zhang2014object} on the same data for comparison (referred to as Had\_PGI). In this experiment, we employed a dog toy with a size of $3\times2~\mathrm{cm}$ as the object. A motorized stage was used to maintain its cyclic translation along the diagonal of the image at an average speed of $6.8~\mathrm{cm/s}$ within the FOV. The DMD operated at 20,000 Hz during the tracking and imaging experiments. We configured a total of 2,000 motion frames and utilized a total of 20,000 patterns, resulting in a measurement time of 1 second. Therefore, the imaging sampling rate of FSI was 61\%, whereas the imaging sampling rates of both Had\_GI and our method were 24.4\%. Figure~\ref{fgr:exp}(b) 
 and Video~\ref{video1} show the motion of the object, reconstructed images, and edges at four different times. With the progression of measurements, an increasing number of Fourier coefficients are acquired, leading to improved image clarity.
\begin{video}
\includegraphics[width=6 cm]{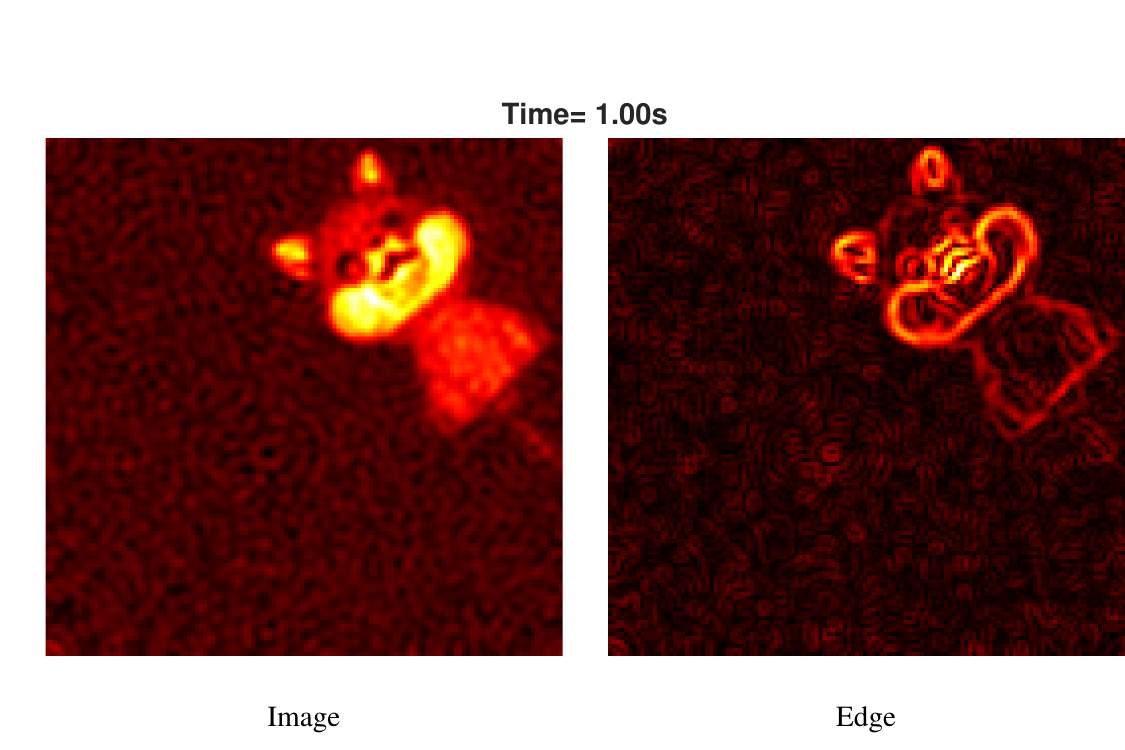}
\caption{Visualization of images and edges reconstructed by our method at different times.
\label{video1}
}
\end{video}

\begin{figure}
 \centering\includegraphics[width=8.6 cm]{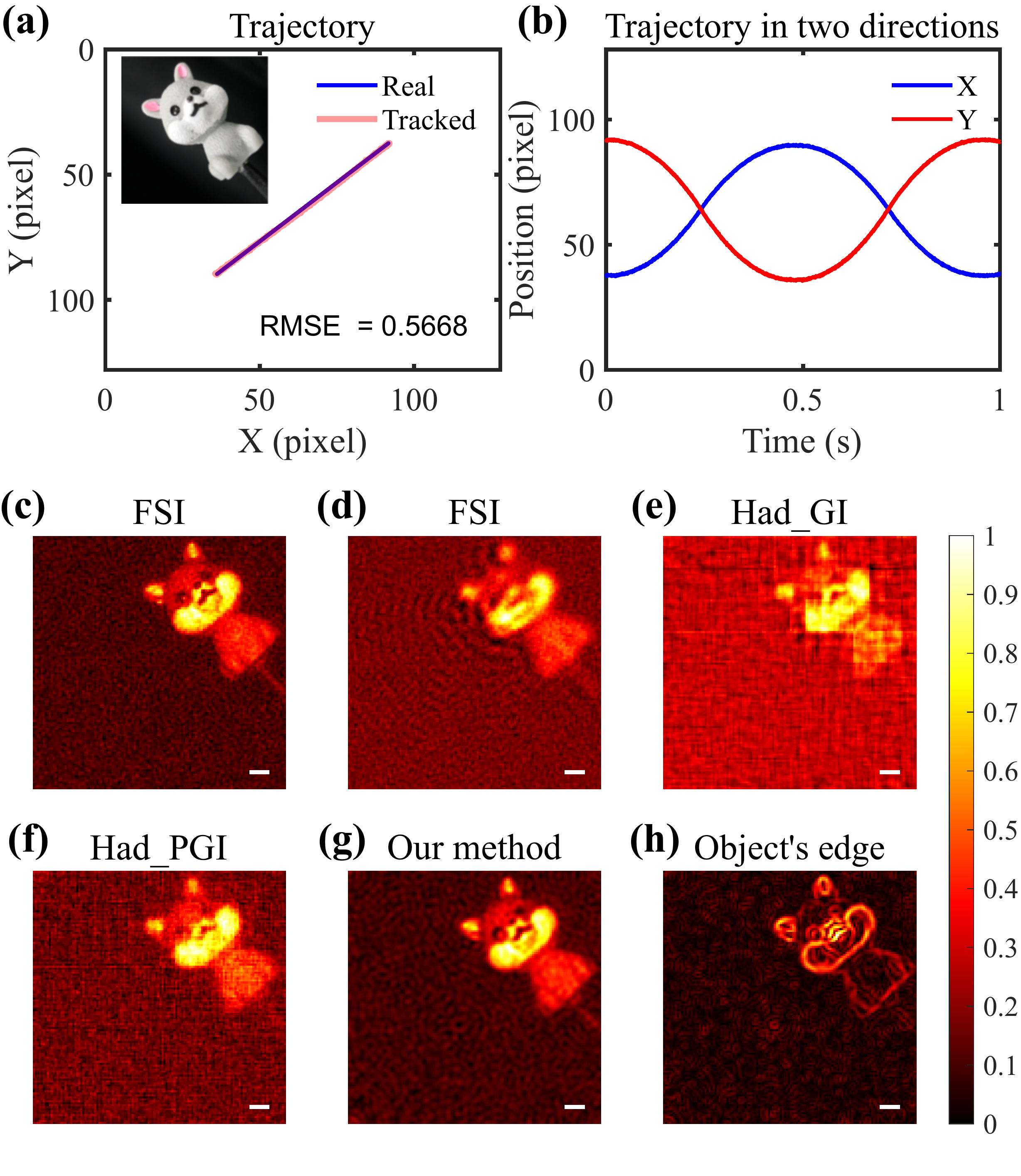}
 \caption{Imaging, edge detection, and tracking results of the moving toy. (a) Real trajectories, tracked trajectories of the moving object within 1 second via our method, and real toy photo taken with a camera. (b) The reconstructed trajectories in two directions. (c) An image of the stationary object obtained via FSI. (d-g) Images of the moving object by FSI, the method using Hadamard patterns and ghost imaging algorithm (Had\_GI), the method using Hadamard patterns and the pseudoinverse ghost imaging algorithm (Had\_PGI), and the proposed method. (h) The edge image restored by our method. All scale bars indicate 5~$\mathrm{mm}$.}
\label{fgr:exp1}
\end{figure}
Figure~\ref{fgr:exp1}(a) shows a real toy photo and a comparison between the tracked trajectory and the actual trajectory within 1 second. Our method exhibited remarkable accuracy, with an RMSE of 0.5668, closely approximating the real trajectory. Moreover, our proposed approach effectively calculates the trajectory in both the $X$ and $Y$ directions, providing a comprehensive view of the acceleration and deceleration of the object within the FOV during the measurement period, as demonstrated in Fig.~\ref{fgr:exp1}(b). When the object is stationary, the image reconstructed via FSI is shown in Fig.~\ref{fgr:exp1}(c). Figures~\ref{fgr:exp1}(d-g) depict the final images obtained by FSI, Had\_GI, Had\_PGI, and our method, respectively. All the experiments were conducted under the same modulation frequency and measurement time. Figure~\ref{fgr:exp1}(h) shows the final edge image of the object restored via our method. 
\begin{table}[b]
\caption{\label{tab:table1} Quantitative evaluation of different methods}
\begin{ruledtabular}
	\begin{tabular}{lccl}
		\textrm{Method}&
		\textrm{PSNR (dB)}&
		\textrm{SSIM}&
		\textrm{Time (s)}\\
             \colrule
		FSI & 18.03 & 0.322 & $2\times10^{-3}$\\
		Had\_GI & 12.64 & 0.201 & $9\times10^{-1}$\\
		Had\_PGI & 18.38 & 0.274 & $3\times10^2$\\
		Our method & 25.13 & 0.558 &$8\times10^{-3}$\\
	\end{tabular}
\end{ruledtabular}
\end{table}
The computation times for FSI, Had\_GI, Had\_PGI, and our proposed method are shown in Tab.~\ref{tab:table1} (tested on a system equipped with Windows 10, an Intel Core i7-9700 CPU clock at 3.0 GHz, and 16 GB of memory). Our method is significantly more computationally efficient than Had\_GI by two orders of magnitude, as it eliminates the need for motion compensation of the imaging patterns. The PSNRs and SSIMs of the reconstructed images were also calculated, as shown in Tab.~\ref{tab:table1}. The image quality achieved through our proposed method surpasses that attained with Had\_GI and Had\_PGI, both in visual appearance and quantitative evaluations. 

\begin{figure}
 \centering\includegraphics[width=8.6 cm]{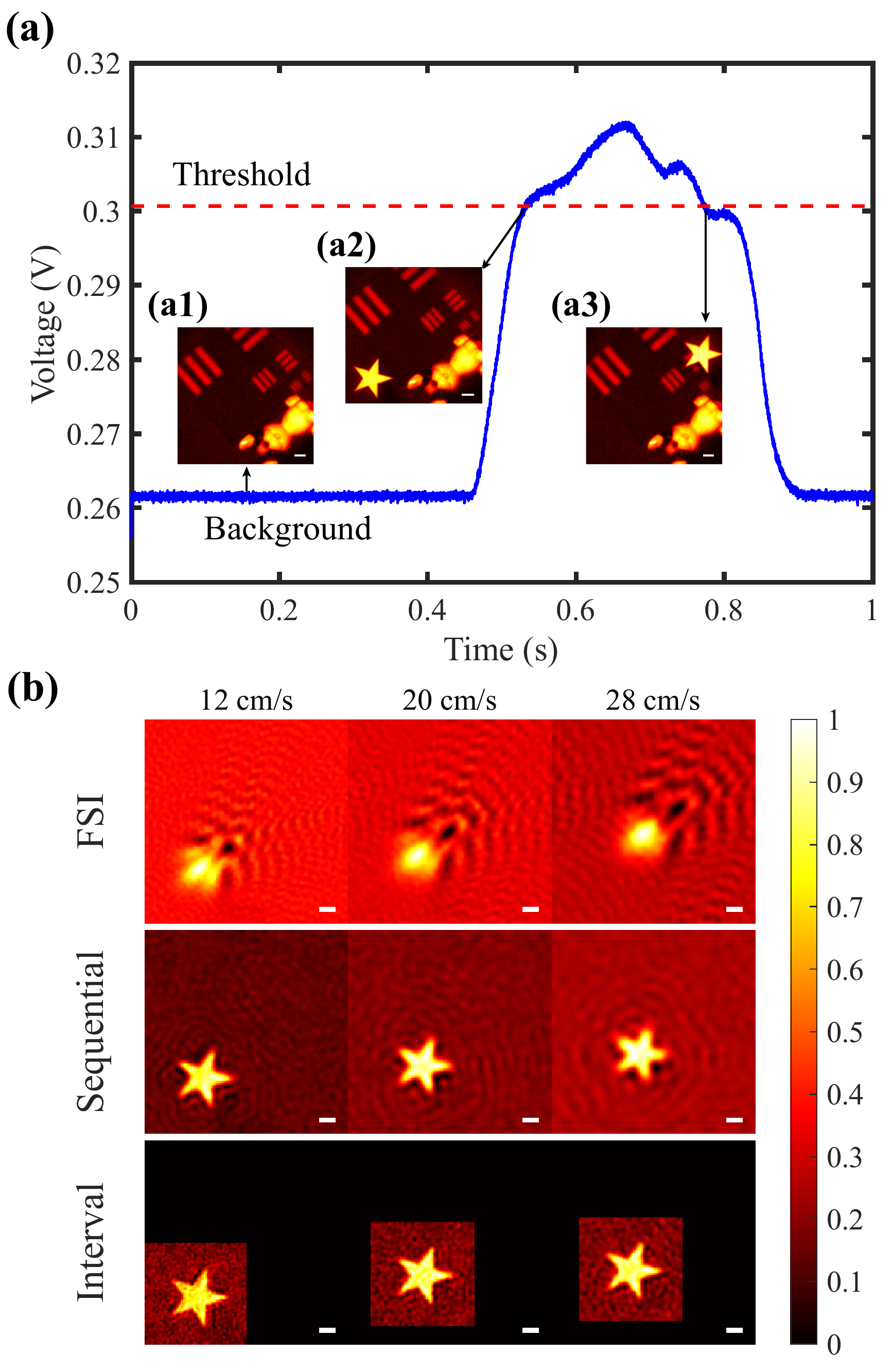}
 \caption{Object monitoring images with the background. (a) The detected readings at various time intervals throughout the monitoring procedure, accompanied by three corresponding images (a1-a3). (b) Images of the object at three distinct motion speeds via FSI, our method with a sequential sampling strategy, and our method with an interval sampling strategy. All scale bars indicate 5~$\mathrm{mm}$.}
 \label{fgr:exp2}
\end{figure}
We designed another experiment involving a dynamic object with a cluttered background to demonstrate the robustness of our approach in handling complex scenarios and the efficiency of our sampling strategy for small moving objects. The moving object was a pentagram constructed from white paper, which traversed the FOV swiftly, resulting in partial overlaps with background objects. Figure~\ref{fgr:exp2}(a) displays the intensity readings captured at various time intervals during this process, as our motorized stage moved at a speed of $20~\mathrm{cm/s}$, alongside three corresponding images. These images represent the background scene, the moment when the object fully enters the FOV, and the instant when the object begins exiting the FOV, respectively. The moment the recorded light intensity surpassed the predefined threshold, it signified the complete entry of the object into the FOV. The program subsequently controlled the DMD to load the modulation pattern sequence until the object exits the FOV. Upon subtracting the previously recorded background data, the trajectory and image of the moving object were restored using the resulting values. The Fourier patterns for imaging were constructed via the three-step phase-shifting method to improve the imaging efficiency, meaning that a total of nine modulation patterns were required for each frame. By utilizing a DMD with a refresh rate of 20,000~$\mathrm{Hz}$, a frame rate of 2222 Hz for tracking positions can be achieved. Figure~\ref{fgr:exp2}(b) and Video~\ref{video2} show the reconstructed images of the object at three distinct motion speeds, utilizing the FSI and our method with two different sampling strategies. These images were restored to their initial modulated positions.
\begin{video}
\includegraphics[width=8.6 cm]{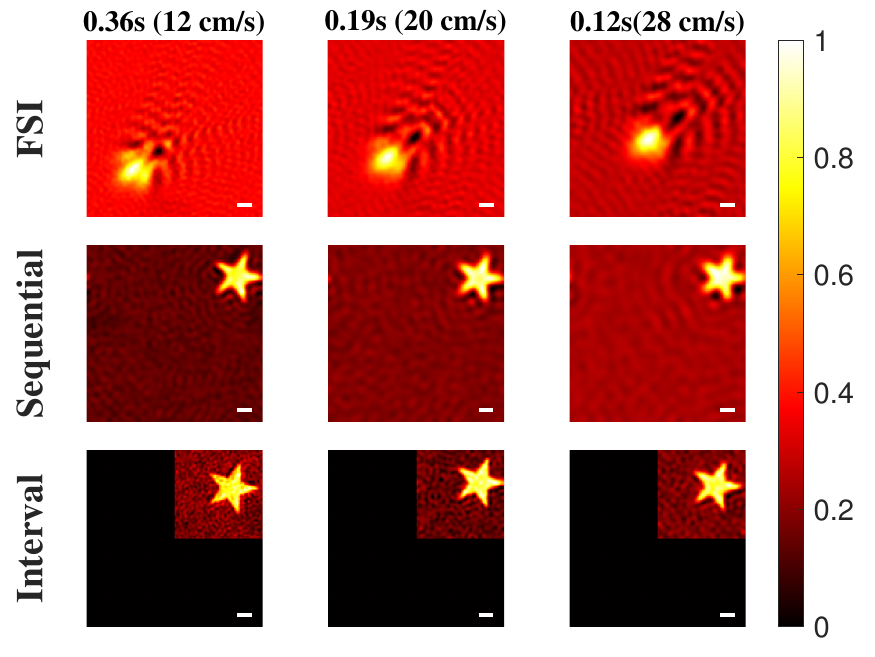}
\caption{Visualization of the reconstructed images of the object
at three diifferent speeds, utilizing the FSI and our
method with two different sampling strategies.
\label{video2}
}
\end{video}
The motorized stage operated at speeds of $12~\mathrm{cm/s}$, $20~\mathrm{cm/s}$, and $28~\mathrm{cm/s}$, equivalent to motion velocities of approximately 276, 460, and 644 pixels per second, respectively. The corresponding dwell times for the object when it was entirely within the FOV were 0.42~s, 0.25~s, and 0.18~s, respectively. Notably, our method outperforms FSI by successfully reconstructing moving object images at all three speeds. Specifically, the interval sampling strategy yields superior results, restoring the finer details of the moving object. As the speed increases, the initial position of the object's reconstructed image gradually shifts in the direction of motion. This phenomenon arises from the delay between detecting the object's entry into the FOV and initiating pattern loading on the DMD. Consequently, higher speeds lead to greater object displacement during this delay period.

\subsection{Discussion}
Owing to the periodic nature of Fourier basis patterns, an object positioned at different locations within the field of view may exhibit the same phase of the measured Fourier coefficient. The higher spatial frequency of the Fourier patterns used for tracking results in more positions with the same Fourier phase, thereby reducing the locating accuracy. In systems with a pixel resolution of $M \times M $, conventional tracking methods that rely on Fourier patterns~\cite{zhang2019image,dan2022motion,li2023single}, typically employ Fourier patterns with a spatial frequency of $2/M$ for tracking, which only provide the relative positions of moving objects. When Fourier patterns with a spatial frequency of 1/M are used for tracking, the Fourier fringe period is $2\pi$, allowing for the unique determination of a position, whereas other spatial frequencies do not. Therefore, our proposed method can achieve absolute position tracking. Moreover, when dealing with small moving objects, such as those sized at $M/2 \times M/2$, the calculated tracking coordinates of the proposed method closely approximate the centroid coordinates of the object. Unlike methods that use Hadamard patterns, our approach enables high-quality object imaging and edge detection with less computational time. Compared with methods aimed at improving the modulation speed~\cite{xu20181000,jiang2020imaging,hahamovich2021single,jiang2021single,kilcullen2022compressed}, the proposed method eliminates the need for post processing the acquired image sequence. The maximum achievable tracking rate of our method is limited by the DMD refresh rate, which is divided by nine, but pattern multiplexing~\cite{li2023single} or complementary detection~\cite{yu2014complementary} can improve the tracking frame rates. Additionally, the reconstruction speed is contingent on the time complexity for the iFFT, enabling real-time imaging capabilities. The background noise in the obtained image can be mitigated further by deep learning-based denoisers~\cite{zhang2018ffdnet,zhang2021plug}. We also acknowledge that our method applies only to translational objects as it relies on the characteristics of the Fourier transform. Thus, the proposed method is unsuitable for imaging objects that undergo rotation or deformation. In addition, if the object’s residence time in the FOV is overly short, it may be difficult to acquire sufficient measurements for reconstructing high-quality images. This limitation can be improved by using an SLM with higher refresh rates.

\section{Conclusion}
In conclusion,  we present a method that uses Fourier patterns for both position information encoding and spatial information encoding. Exploiting the unique characteristics of the Fourier transform, the proposed method achieves improved image quality and edge detection with low time complexity and minimal memory consumption of a moving object. Additionally, our selective use of Fourier patterns with specific spatial frequencies addresses the limitations of the Fourier pattern-based tracking methods, enabling precise object location determination. Moreover, we present an interval sampling strategy specifically designed for small moving objects, which greatly reduces the required dwell time. The potential applications of the proposed method span various fields where the tracking and imaging of high-speed moving objects in real-time are essential.

\section*{}
\begin{acknowledgments}
This work was supported by the Beĳing Institute of Technology Research Fund Program for Young Scholars(Grant No.20212012).
\end{acknowledgments}

\bibliography{apstemplate}

\end{document}